## Chapter 1

# Redispersibility in magnetorheological fluids and its relevance for MRF formulations

*Sara R. Gomes de Sousa[1] and Antonio J. Faria Bombard[1]\**

Very important research, with both theoretical and experimental results in the advancement of physical models that explain the MR effect, try to keep the MRF formulation as simple as possible, usually with only two components: a magnetic dispersed phase and a carrier liquid. Many patents of MRF include three or four components, such as some surfactant and thixotropic additives. In order to formulate a good and reliable MRF for different applications such as MR shock absorbers, clutches, brakes, the MRF redispersibility is a challenge, but mandatory, key property for out of lab real world applications. This chapter, we shows how to measure and evaluate the MRF redispersibility.

## 1.1 Introduction

According to the Merriam-Webster Dictionary, dispersibility is defined as: "*the quality or state of being dispersible*". Dispersible is an adjective related to the transitive verb: to disperse, which among other meanings, has the following meaning in the context of chemistry: "*to distribute (something, such as fine particles) more or less evenly throughout a medium*". Therefore, one can understand redispersibility as the ability of a dispersion or suspension (in the colloidal physical-chemical context) to be redispersed.

In this chapter, we will be more interested in how easy or difficult it is to redisperse a sample of magnetorheological fluid (MRF) after some time (days, weeks, months, years…) under rest; usually, under normal gravity – 1 'g'. Sometimes, one can be interested in measuring the redispersibility of MRF after subjecting it to centrifugal force. In this case, one must specify how many 'g`s were applied to the sample, and how much time the sample was centrifuged as it directly affects the sediment.

It is quite common, especially when beginners start to work in magnetorheology, to be concerned with the MRF "stability."It is important to understand the difference between "sedimentation stability", related to Stokes' Law and the large density mismatch of the dispersed phase (usually dense, magnetic solid particles, with a size range of ∼ 0.1 – 10 μm, as iron microparticles) and the liquid-dispersing medium (whose densities ranges 786 – 1,890 kg/m³ at temperatures ∼ 25°C). By far, the most common magnetic material investigated and employed to formulate MRF is carbonyl iron powder. Its density is ∼ 7,860 kg/m³.

---

[1]Federal University of Itajubá, Physics & Chemistry Institute, Itajubá MG, BRAZIL
\* corresponding author: bombard@unifei.edu.br

On the one hand, sometimes one wants to keep the MRF as simple as possible to get theoretical, physical models to explain and fit experimental results well. These MRFwere prepared only with oil (perhaps other liquid) and the magnetic phase. Usually, these MRF samples settle very fast, and the only sedimentation retarders are the parameters of Stokes' Law (liquid viscosity, particle size, density phasesmismatch).

On the other hand, even those commercial MRF well formulated, with proprietary additives (dispersing, thixotropic, anti-wear, etc.) and that are sold by large industries such as LORD Inc. or BASF SE, are prone to show phase separation. Additionally, anyone who bought commercial MRF or prepared their own MRF knows well that this phase separation is just a matter of time.

However, much more serious than the old problem of settling (or sedimentation and phase separation), intrinsic to MRF, is the problem of colloidal stability. Related "cousins" of MRF, the ferrofluids, do not show phase separation since they are true colloids; their particles are in nanometer size range, usually 1 – 10 nm. Ferrofluid's, if well chemically prepared, are very stable. Besides not settling, surfactants and Brownian motion keep their particles apart, in such a way that they are not subject to coagulation.

This phenomenon, <u>coagulation</u>, in the colloids context, is the more serious problem when formulating a good MRF. Without dispersing additives, the only thing that could hinder the iron particles to coagulate is if the van der Waals forces are null. As these forces are always omnipresent in almost any dispersion, the only way to cancel them would be to match the Hamaker constant in the magnetic phase (iron powder or other) with the Hamaker constant of the liquid phase [1]. (Israelachvili 2011). Besides van der Waals attractive forces, magnetic materials also can aggregate due to remnant magnetization, and affects the redispersibility of MRF [2] (Phulé et al 1999).

Several approaches were tested, including coating the iron core micro-particles with thin, nanometric shells of different polymers. This can be a solution, although turning the carbonyl iron particles - expensive themselves – is even more expensive.

Another possible way which is easier and more time and money efficient, is to test (even by trial-and-error) different dispersing additives and find one that modifies the surface of the magnetic phase, matching the "physical-chemical" nature of the dispersed phase with the carrier liquid Hamaker constant. Following one of the recipes described in the US patent # 5,645,752: "Thixotropic Magnetorheological Materials" assigned to LORD Corp., invented by Weiss *et al.* (1997) [3], one of the authors of this chapter prepared a lot of MRF using carbonyl iron powder, a silicone oil (different branch of that in patent, but also silicone oil), and the same dispersing additive: "SILWET L-7500" (Witco / Momentive chemicals) described in the patent. This lot of MRF was prepared in 1999. After years under rest, settling and phase separation of iron powder were observed several times since then. However, just shaking the bottle containing the MRF by hand was enough to add energy to redisperse the entire dispersion completely. We think that it is a good example of this approach.

Now we will describe how to measure and evaluate the redispersibility of an MRF sample.

### 1.1.1 Methodology of the redispersibility test

Although one can always use a simple spatula, according to ASTM D869 – 85, or the so called "Instinctive hardness test" (Portillo and Iglesias 2017) [4], we will describe here the approach first employed to test the redispersibility of an MRF (to the best of our knowledge) by Kieburg *et al.* 2006 [5].

This is a modified, rheometer penetration test version of the ASTM D869 – 85: "*Standard Test Method for Evaluating Degree of Settling of Paint*" (Reapproved 2015) [6]. Table 1 below is an adapted short, partial reproduction of the recommended Rating in ASTM D869 – 85. More detailed, full rating, refer to ASTM D869 – 85.

*Table 1 Hand drop Spatula with 45 ± 1 g, 125 mm length x 20 mm width into a can with 0.5 L of paint, after six months (or time agreed). – ASTM D869-85.*

| 1 | **Paint condition** |
|---|---|
| 10 | Perfect suspension. No change from the original condition. |
| 8 | A feel of settling and slight deposit. No significant resistance to sidewise movement of spatula. |
| 6 | Definite cake of settled pigment. Definite resistance to sidewise movement of spatula. |
| 4 | Difficult to move spatula through cake sidewise and slight edgewise resistance. |
| 2 | If spatula has been forced through cake, it is very difficult to move sidewise. |
| 0 | Very firm cake that cannot be reincorporated with the liquid to form a smooth paint by stirring manually. |

\* Intermediate conditions between above, are given the appropriate odd number.

The above table is just to show how subjective the "hand spatula" method can be in quantifying the redispersibility. However, if the other way is not available, using a steel spatula and human tactile feeling is a good tool to compare (at least qualitatively) MRF formulations after sedimentation.

Instead, a subjective, only qualitative rating evaluation based on a tactile feeling of a human hand, the normal force, capacitive sensor built-in some commercial rheometers, is employed to measure how easy/difficult is to penetrate the iron powder cake sediment. The test can be run in an MRF sample container like a test tube (15 mL volume, or with ~ 100 mm height). If available, a rheometer equipped with Normal Force sensor, vane rotor and vertical downward movement with controlled, constant speed, is preferable.

In order to accomplish the test with a rheometer, one test tube containing the MRF sample is left under rest, to settle under normal gravity. After the time scheduled (24 hours, 1 week, 1 month, etc.), the test tube is put under the center aligned with the rheometer head, and a penetration test is run. One can employ a needle or a Vane tool, or even a steel blade (e.g.: 50 mm height x 10 mm width x 1 mm thickness) attached to the rheometer coupling, and the head is moved downward at constant speed. Usually the movement speed in set on the software to 1 mm/s for instance, and taking advantage of the rheometer normal force readings and register points each 0.05 second (or 50 μm penetration). In addition, an event control can be programmed in the rheometer software, so that the test is automatically interrupted if the normal force reaches the maximum that the capacitive sensor of the rheometer supports (50 N in our case).

Another possibility, as demonstrated by Iglesias *et al* 2012 [7];Portillo and Iglesias 2017 [4], is to use an analytical scale and a mechanism employing a stepper motor to move the tube containing MRF to be tested under the load cell of electronic scale.

Figure 1 shows a steel blade tool attached to the rheometer head (Anton Paar MCR 301) and a tube containing the MRF sample in position to run the penetration redispersibility test. The head started at the top position (arbitrary) and is moved downward, penetrating the MRF and its iron powder content (or any other magnetic solid sediment). All the tests here were performed with MRF sedimenting at room temperature, over months or even years. But the variation of temperature oscillated between 15 - 35 °C considering winter and summer in place of the experiments.

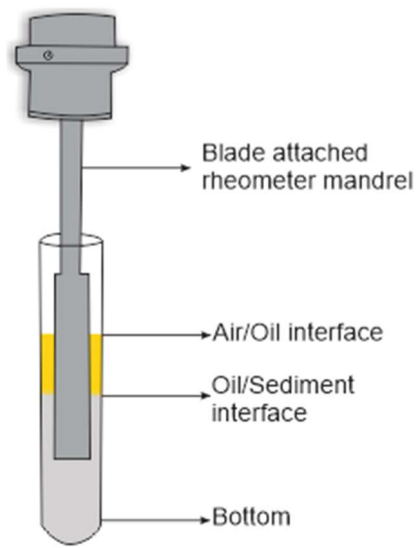

*Figure 1 The illustration of the steel blade tool attached to the rheometer head and a tube containing the MRF sample in position to run the penetration redispersibility test. A vane rotor also can be used.*

## 1.2 Results and Discussion

### *1.1.2 Effect of sedimentation time*

The next figures will present redispersibility test curves in which normal force (N) is a function of the depth (mm). Although the X axis can also be presented as the gap, simple subtraction can be used to invert it. Instead of the value of the gap (distance between the lower tip of blade or vane) to the tube bottom, one can show the depth as the distance the tool penetrates the sediment. We prefer this way to present our results, but both (gap or depth) give the same information.

Figure 2 shows the results for the same MRF, with 80% by weight of carbonyl iron powder and additives. These were tested after one week, one month, five months and ten months of sedimentation, in normal gravity. The normal force resolution of the capacitance sensor of the rheometer used for the tests is 1mN.

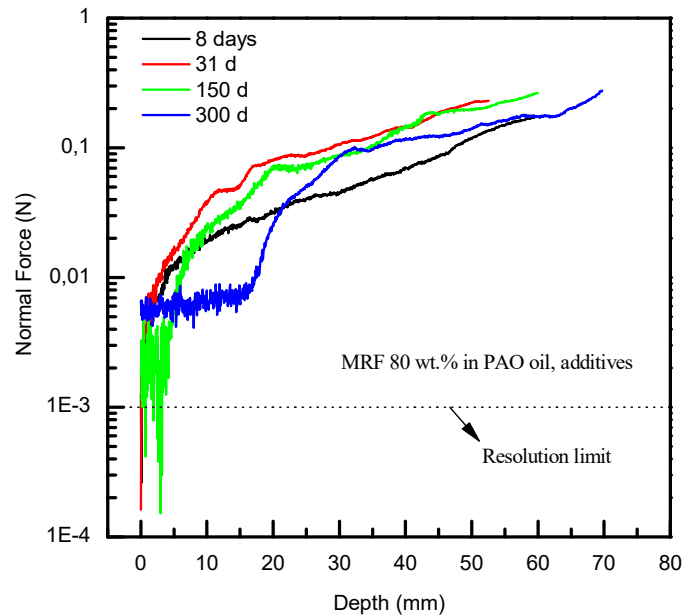

*Figure 2. The normal force measured as a function of depth after the penetration test. MRF prepared with 80 wt.% carbonyl iron powder in PAO oil. Curves tested after one week, one month, five months and ten months of settling.*

In Figure 2, it is observed that even after almost a year under rest, the sediment did not exceed 0.3 N of peak in the normal force. In addition to the fact that the curves did not change much from one week to one month or ten months, they indicate that this formulation can be considered 'easy' to redisperse or that it has good redispersibility.

Integrating the area under the $normal force x depth$ curves gives the mechanical work (energy) exerted by the blade when it penetrates the sediment. The higher the value of the work shows that it is more compact and hard to penetrate. However, it must be considered that being a mathematical product: $w = Fxd$, we should not only analyze the value of the work.

There may be cases where only 1 mm of penetration the normal force reaches a peak of 40, 50 N (50 N is the operating limit of the MCR-301 rheometer).In this case even if the work results in a value of 40 mJ, it is a result that indicates the formation of a very hard sediment (*hard cake*). Thus an MRF of redispersibility becomes difficult or impossible after a certain point. In other words, it is considered a poorly formulated MRF.

Accordingly, we have to consider both the $normal force x depth$ curves and the $work x depth$ curves in the evaluation of redispersibility. Figure 3 below corresponds to the integration of the areas under the curves of Figure 2, above.

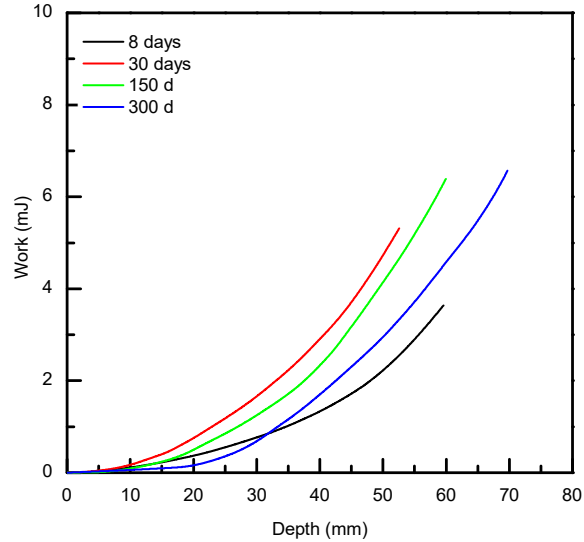

*Figure 3. Work (mechanical energy) spent for blade penetration inside the sediment in MRF test tube, after different settling times. Curves calculated by integration of the curves in figure 2. Same experimental conditions.*

It can be observed in Figure 3 that the work, even after ten months of sedimentation, was below 10 mJ confirming that this can be considered a well formulated MRF. Even after 300 days, an MRF remains 'easy' to redisperse. The results of Figure 3 are summarized in table 2.

*Table 2. Work of 80 wt.% MRF in PAO oil, figure 2 and 3.*

| Time (days) | Max Normal Force (N) | Depth (mm) | Work (mJ) |
|---|---|---|---|
| 8 | 0.15 | 60 | 3.6 |
| 30 | 0.20 | 52.6 | 5.3 |
| 150 | 0.25 | 60 | 6.4 |
| 300 | 0.28 | 70 | 6.6 |

Figures 2 and 3 and Table 2 are examples of good MRF formulations in terms of redispersibility since the maximum force for blade penetration into the sediment, even after 300 days of settling under rest, was below 0.3 N, and the total work each test always below 10 mJ for 50 mm (or more) penetration.

It is important to understand that, although the workas mechanical energy spent by the blade to penetrate the solid powder sedimentcan be useful in comparing different MRF formulations,this is merely secondary information and Normal force peak together with the depth penetration is much more significant.

For instance, let us to consider the hypothetical examples: A MRF 'A' which presents maximum normal force of 0.5N, a depth of 50 mm and penetration work of 25 mJ. Then, a second MRF 'B' with the same working value of 25 mJthat reached 25N of normal force with only 1 mm of depth.

It would be possible to think that the two formulations of MRF 'A' and 'B' are good considering only the parameter 'work' since both presented the value of 25mJ. However, during the penetration test the MRF 'B' had a depth of 1mm and a normal force of 25N indicating that there is a formation of hard cake, and therefore, it is a bad MRF formulation. Different the MRF 'B', the MRF 'A' reached a depth of 50mm reaching only 0.5N of normal force that could indicate a good MRF formulation.

As an example of the description above, we may can consider the example of Figure 4. Below and in sequence,Table 3 shows the results of the curves shown in Figure 4 in an easy way.

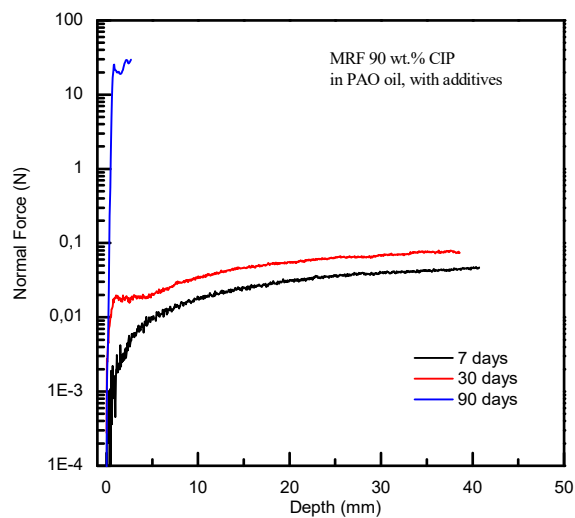

*Figure 4. Redispersibility curves for an MRF with 90 wt.% CIP in PAO oil, with additives, sedimentation under normal gravity, at rest. Black curve measured at seven days, red curve at one month, and blue curve at three months.*

*Table 3. Results extracted from Figure 4 where normal force is shown along with depth and work after a certain time (7 day, 30 days and 90 days).*

| Time (days) | Max Normal Force (N) | Depth (mm) | Work (mJ) |
|---|---|---|---|
| 7 | 0.05 | 40.7 | 1.15 |
| 30 | 0.075 | 38.6 | 1.94 |
| 90 | 29.5 | 2.68 | 48.8 |

The redispersibility curve after one week shows that the formulation of this MRF appears to be excellent since the normal force did not exceed 0.05 N at a depth of 40 mm. The same could be said after a month of sedimentation with the force below 0.10 N at practically the same depth. After three months of rest, the test indicated a force of 30 N at only 2.7 mm depth, clearly demonstrating the formation of a hard cake. Therefore, the MRF in question does not have good redispersibility, although it may appear in the first 30 days.

In this case, the work and the normal force peak agree, demonstrating that at times before one month the formulation seems to be acceptable, but after three months, it changed and showed hard cake sediment.

### 1.1.3    Effect of additives in MRF formulation

To make it clear that raw data of normal force vs. penetration depth in the redispersibility test is more important than just the mechanical working value, the example below (Figure 5)illustrates 3 MRF formulations that have all been assayed after one month of sedimentation. The iron powder and the concentration was the same in the 3 formulations. The only difference between the 3 MRFs was the dispersant additive employed in each. The test was programmed (via event control in the software) to abort the test if the force reached 30 N. Table 4, below Figure 5,summarizes the values of Normal force, depth and work for each MRF tested.

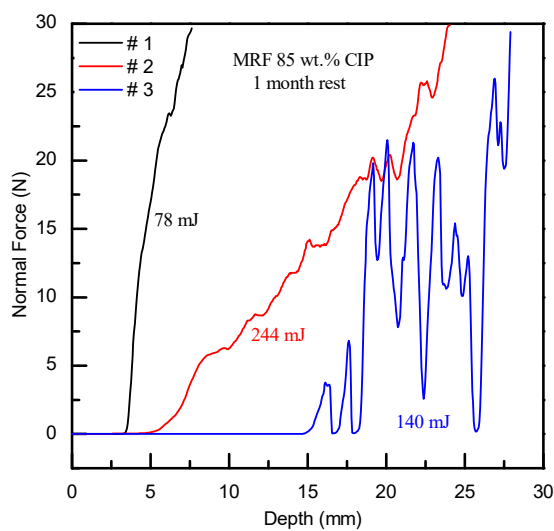

*Figure 5. Redispersibility curves for MRF with 85 wt.% CIP in PAO oil, with three different dispersing additives, all tested after one month of sedimentation under normal gravity, at rest. The numbers 78, 244 and 140 mJ are the corresponding values of work, by integration of the areas below the curves.*

*Table 4. Results extracted from Figure 5 which shows normal force, depth and work after one month rest for the different additives.*

| MRF with additive | Normal Force peak (N) | Maximum Depth * (mm) | Work (mJ) |
|---|---|---|---|
| # 1 | 29.67 | 4.27 | 78 |
| # 2 | 29.9 | 20.0 | 244 |
| # 3 | 27.9 | 13.56 | 140 |

* The depth here considered as the distance traveled by the blade between the first layer of iron powder sediment (mud line) and the 30 N limit reach.

If one takes into account only the work spent on the penetration of the blade, one might think that additive # 1 is the best since this formulation had a lower working value (78 mJ). However, since the experiment was limited to a normal force of a maximum of 30 N, all formulations have a peak force value of ∼ 30 N. In this case, what is more relevant is the distance that the blade has been able to penetrate. The smaller the distance - independent of the area under the curve (ie.: the mechanical work worn) - the more compact and harder and also more difficult it is to redisperse the sediment formed in the sample composition.

Therefore, in the case of the samples of Figure 5 and Table 4, Additive #1 was the worst of the three, Additive #3 was regular, and Additive #2 was the best since it presented the longest distance of penetration which also indicates a less compacted sediment.

This shows that the value of the parameter 'work' is relative and must be used with caution, and always in conjunction with the force-depth curves. In this case, none of the 3 formulations showed a good result after one month. But it is a case that illustrates just how mechanical work – alone – should not be taken as a criterion for evaluating the redispersibility of an FMR well, at least not in the penetration test. Thus, the parameter 'work' is only relevant when comparing the same depth of sediment.

To demonstrate this, see Figure 6 below. In this case, three completely different additives from those of Figure 5 were used in the preparation of new FMR formulations. After one year of rest and sedimentation, the fluids were analyzed in the redispersibility test. All were subjected to the same penetration distance while keeping the *event control* when the depth reached 1 mm before the bottom of the container tube. The force applied was limited by 50 N, which did not occur for any of the three formulations.

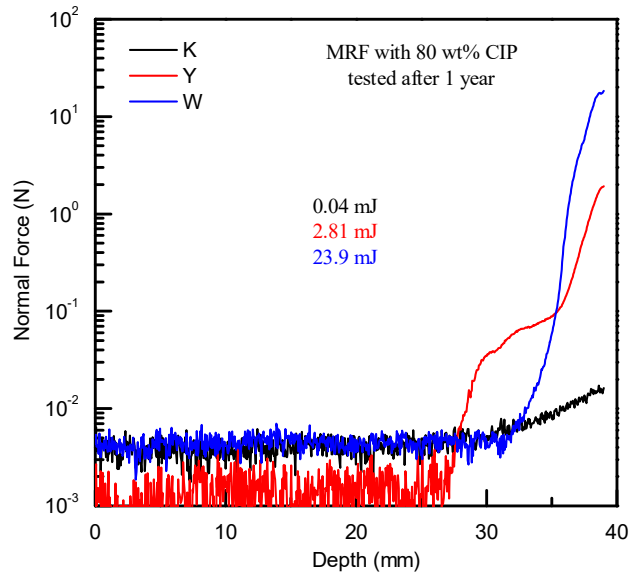

*Figure 6. Redispersibility curves for MRF with 80 wt.% CIP in PAO oil, with three different dispersing additives, all tested after one year of sedimentation under normal gravity at rest. The numbers in front the additive code letters (K, Y and W) are the corresponding values of work by integration of the areas below the curves.*

In Figure 6, we can see that in this case, the additive 'K' was the best of the three additives used, with an excellent redispersibility. Considering that after one year the work did not get to 0.05mJ and the peak-force 0.018N, Additive 'Y' was acceptable with a work of ~3mJ and a peak-force of 1.93N. However, Additive 'W' was the worst with almost 24mJ of work and a peak-force of 18N after the same amount of time.

The figures 7 and 8, shows the redispersibility test curves for MRF formulated with 85 wt.% CIP, but with two different dispersing additives (D-1 and D-2).

The table 5 summarizes the work (as mechanical energy), spent while the blade penetrates the sediment at different times of settling.

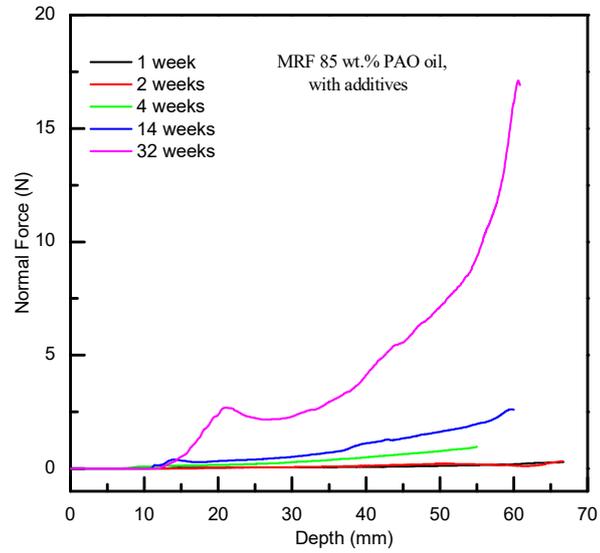

*Figure 7. Redispersibility test curves for an MRF sample with 85 wt.% ( ~ 36 vol.%) CIP in PAO oil, with additives dispersing 'D-1' and thixotropic 'T' after different times of settling under rest, normal gravity.*

*Table 5. Work in Figure 7*

| Weeks | Work (mJ) |
|-------|-----------|
| 1 | 5.5 |
| 2 | 6.2 |
| 4 | 18 |
| 14 | 48 |
| 32 | 232 |

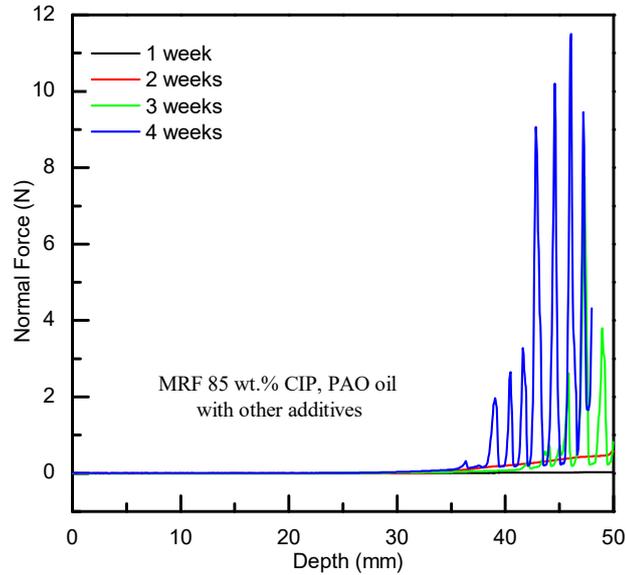

*Figure 8. Redispersibility test curves for an MRF sample with 85 wt.% (∼ 36 vol.%) CIP in PAO oil, with additives dispersing 'D-2' and thixotropic 'T', after different times of settling under rest, normal gravity. Table 6 resume the work values associated with Force x depth curves in figure 8.*

*Table 6. Work in Figure 8*

| Weeks | Work (mJ) |
|-------|-----------|
| 1 | 0.4 |
| 2 | 4.3 |
| 3 | 9.1 |
| 4 | 24.6 |

As can be seen from Figure 7 and Table 5, it is not so difficult to formulate an MRF whose redispersibility, as evaluated in short time intervals, is good and acceptable. After three months or longer, it is not easy to get a low value of normal force with its associated work.

On the other hand, employing the same CIP and poly(alpha-olefine) oil, at same volume fraction, but with different dispersing additive 'D-2', the redispersibility behavior can change completely according to Figure 8.

Again, if only the redispersibility work valuesare compared, one could think that Additive D-2 is goodbecause the total work after four weeks of resting is relatively low. However, from Figure 8, the pattern of peaks and valleys, which can occur with some formulations, is still poor after only three weeks, reaching 10 N and more which is a bad result. This is another example that one must be careful to not evaluate only the values of work, but also the peak of normal force measured in the redispersibility test.

### 1.1.4   Effect of centrifuging the MRF

The figures 9 and 10 shows respectively, normal force x depth and work x depth, for a commercial MRF sample, after 1 week and 18 weeks, tested under normal gravity. The same MRF, but after centrifuged under 2,000 'g' by 15 min, was also tested. Result is presented in figure 11.

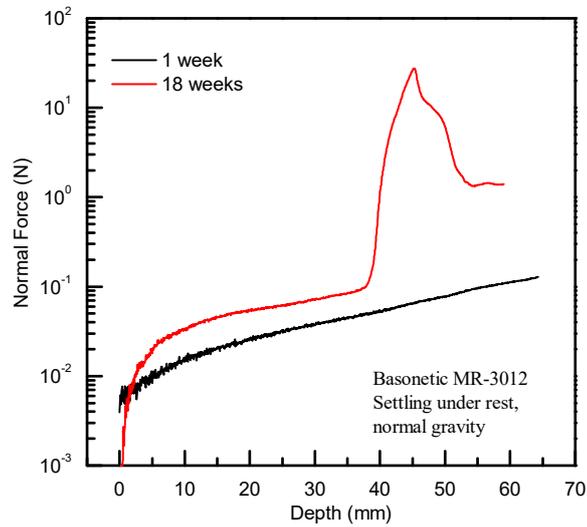

*Figure 9. Redispersibility test curves, for a sample of commercial MRF (Basonetic MR-3012, BASF SE). Settling under rest (normal gravity), after one week (black curve) and 18 weeks (red curve).*

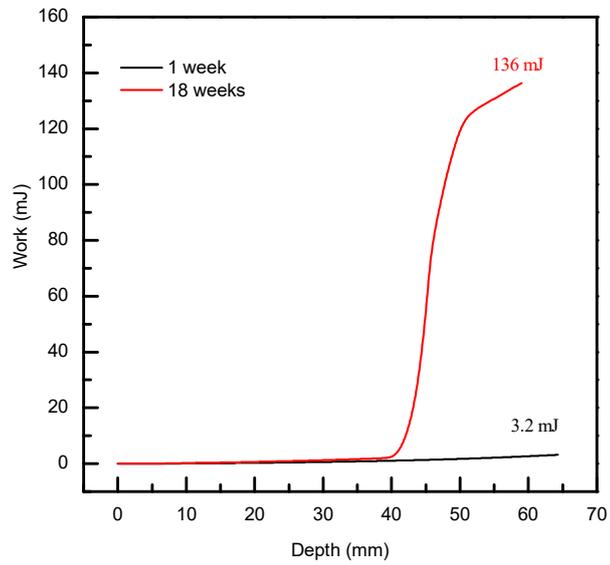

*Figure 10. Work (calculated by integration, of the area under curves of Figure 9), spent by blade penetration in the sediment of MRF Basonetic 3012.*

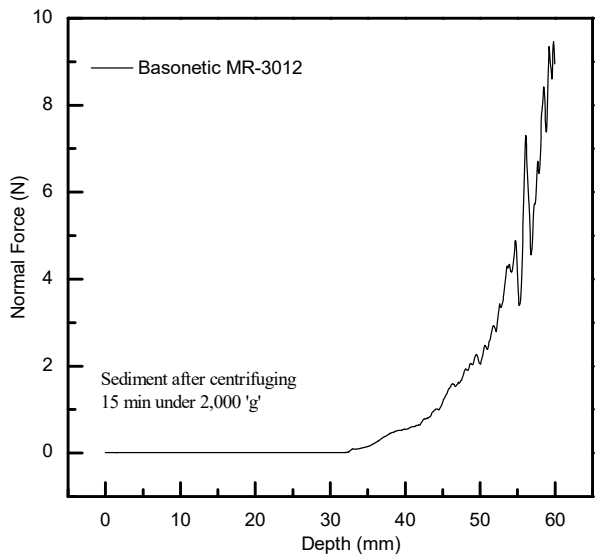

*Figure 11. Normal force x depth, for a sample of MRF Basonetic 3012, after centrifuging it inside a test tube under 2,000 'g,' for 15 minutes. The first 30 – 35 mm of depth, the blade was moving downward in the air, or just starting to penetrate the*

*liquid supernatant. Note the rapid increase of the registered normal force, at depth > 40 mm.*

The figure 12 shows the effect of centrifuging on three different MRF: one commercial (BASF MR-3012), and two of our MRF formulations. One was prepared with 90 wt.% one size CIP. The other sample is a blend of two sizes.

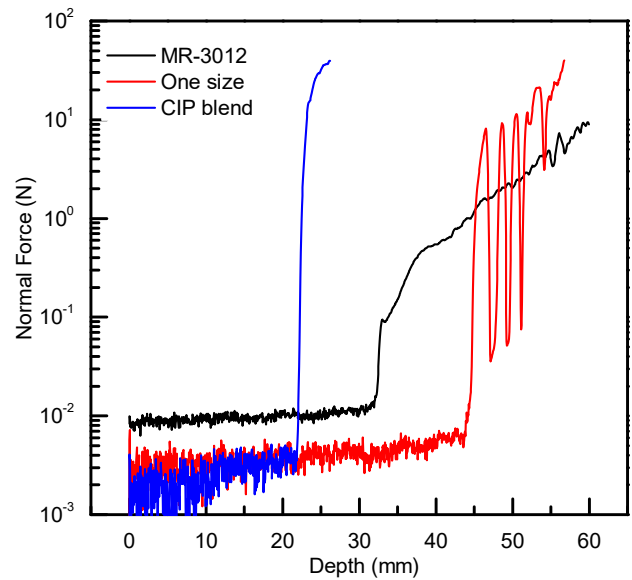

*Figure 12. Redispersibility test curves for different MRF samples, centrifuged under 2,000 'g' by 15 minutes. Comparison between a commercial MRF (Basonetic MR-3012, black curve) and two other MRF with 90 wt.% CIP: prepared with only one CIP size (red curve) and a blend of two sizes (blue curve).*

In Figure 12, the region with 'noise' corresponds to the region where the blade attached to the rheometer is moving downward, but in the air or inside the upper liquid supernatant of MRF sample. Note that for the MRF with blend sizes, the blade penetrated only 4.3 mm, and reached 40 N of normal force very quickly, an indication that its sediment was very hard to penetrate. Therefore, a bad result. The sediment for the MRF prepared with only one size also reached 40 N, but the blade penetrated ~ 13 mm, and with a succession of increase-and-decrease of thenormal force. Thissuggested that the sediment was formed in layers. The best result, in this case, was obtained with the commercial MRF sample. In general, mixing sizes of iron powders can improve the redispersibility (Teodoro and Bombard) [8], but not always, as figure 12 illustrate.

### 1.1.5 Long time settling (one year) redispersibility

A well formulated MRF, like that shown in thefigure13 below, resulted in theredispersibility test with Normal forces lower than 0.1 N, and work around 1 mJ, for ~ 20 mm blade penetration into iron sediment after one month of settling under rest. Even after more than one year of settling under rest, the normal force reached less than 0.13 N and work of ~ 1.8 mJ.

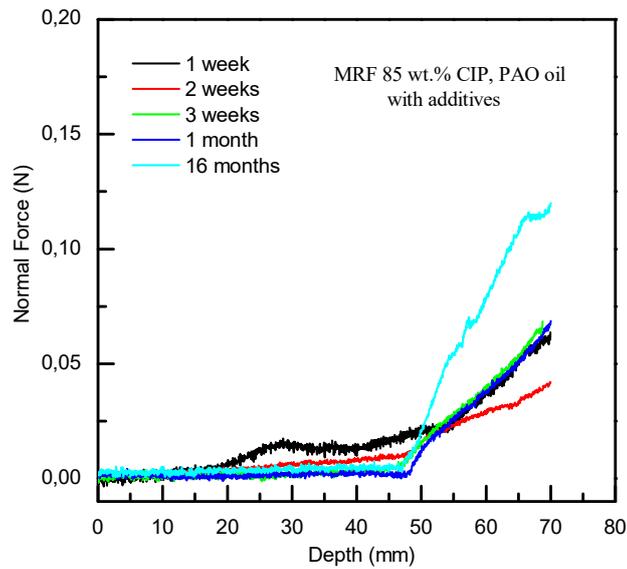

*Figure 13. A well formulated MRF, after one month of sedimentation,*
*shows less than 0.1 N of the normal force.*

*A priori* it can be apparent that to get a good result in redispersibility test, when one is testing several different dispersing additives for iron powder and seeking to formulate a good MRF. It is simple, considering only short (7 days) times of settling under rest. This is shown in thefigure14, for MRF all formulated with 85 wt.% carbonyl iron powder, but different additives, after one week of sedimentation.

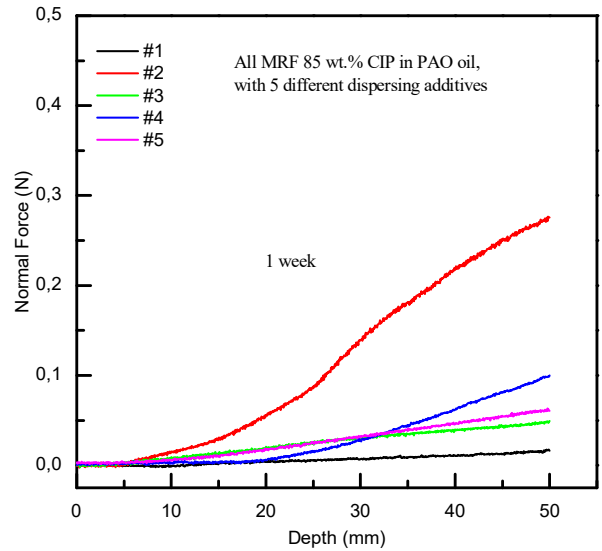

*Figure 14. Redispersibility test curves for five formulations of MRF, all with 85 wt.% CIP in same PAO oil, but with different dispersing additives. Settling under rest (normal gravity), after one week.*

### 1.1.6   Redispersibility of MRF with 48 vol.%

The figures 15 – 17, shows the redispersibility test results for MRF with high volume concentration ~ 48 vol.% (or 90 wt.%) of iron powder, after three months under rest. It is evident the relevance of the additives employed.

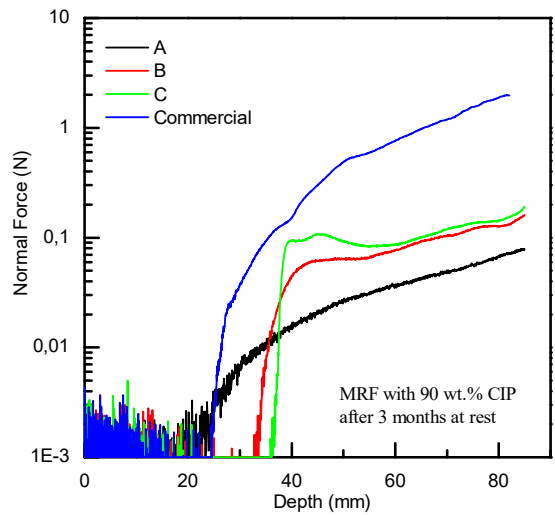

*Figure 15. Redispersibility test curves of four formulations of MRF, all with 90 wt.% CIP, but with different dispersing additives. Settling under rest (normal gravity), after three months. One line (blue) is a commercial MRF sample, Basonetic 3012.*

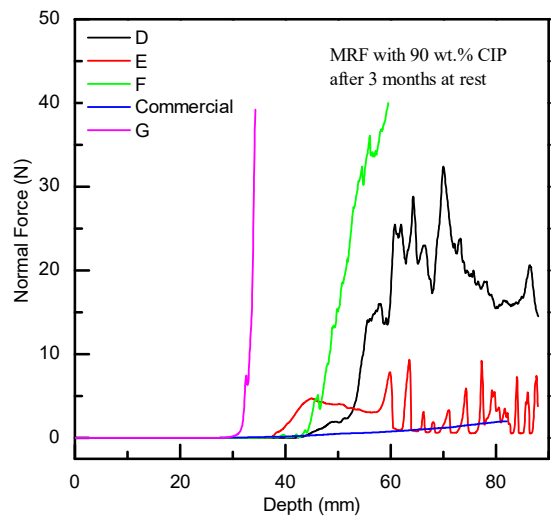

*Figure 16. Redispersibility test curves for five formulations of MRF, all with 90 wt.% CIP, but with different dispersing additives. Settling under rest (normal gravity), after three months. One line (blue) is a commercial MRF sample, Basonetic 3012.*

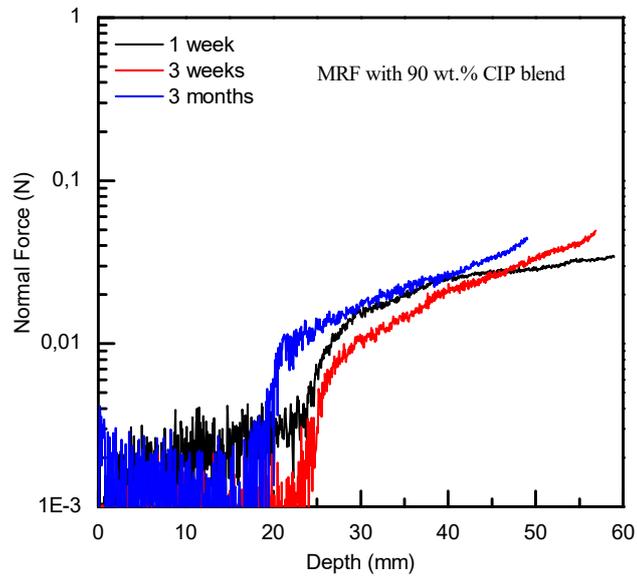

*Figure 17. Redispersibility test curves for one MRF formulations with 90 wt.% CIP ( ~ 48 vol.%) size blend. Settling under rest (normal gravity), after one week, three weeks and three months.*

**Conclusions**

Redispersibility in magnetorheological fluids is determined to be 'good' when after a long period of time the MR fluid does not require much work to disperse while maintaining a penetration depth equal or close to the initial one.

It is also possible to analyze different redispersibility using different dispersing additives where each one will act and respond in a different way in obtaining the curves of normal force vs. depth and consequently in the obtained values of work.

We should be careful when initially analyzing an MR fluid as soon as it has been prepared because it can be very good and after a short period of time it starts to show 'hard cake' as a 'good' fluid.

In other words, the redispersibility of an MR fluid, in particular, cannot be analyzed and concluded to be 'good' with a single analysis of response work, but rather as a junction of the depth and normal force curves after long periods of rest time.

**Acknowledgments**

Mrs. Sara R. G de Sousa acknowledges CAPES by her PhD student fellowship. AJFB acknowledges FULBRIGHT (US Dept. of State, Bureau of Educational and Cultural Affairs) the financial support during his 3 months research time at University of Wisconsin – Madison. We also thank FAPEMIG Grant APQ-01824-17.